\documentclass[12pt,tightenlines,eqsecnum,floats,shownopacs,nofootinbib,amsmath,amssymb,aps,prd]{revtex4}

\usepackage{amssymb}
\usepackage{amsmath,amssymb,amsfonts,amsthm}
\usepackage{graphicx}

\def\be{\nopagebreak[3]\begin{equation}}
\def\ee{\end{equation}}
\def\ba{\nopagebreak[3]\begin{eqnarray}}
\def\ea{\end{eqnarray}}

\def\lp{\ell_{\rm Pl}}

\def\mpl{m_{\rm Pl}}

\def\rcr{\rho_{\rm max}}

\def\LQC{\rm LQC}

\def\B{{\rm B}}

\def\bu{$\bullet\,\,$}

\def\t{\tilde}
\def\h{\hat}

\def\tr{\rm Trun}

\def\x{\vec{x}}

\def\vk{\vec{k}}
\def\R{\mathcal{R}}

\def\g{\mathfrak{A}}

\def\T{\mathcal{T}}
\def\ps{\Gamma}

\def\H{\mathcal{H}}

\def\R{\mathcal{R}}
\def\Q{\mathcal{Q}}

\begin{document}

\title{Loop Quantum Gravity and the The Planck Regime of Cosmology}
\author{Abhay Ashtekar}
\affiliation{Institute for Gravitation and the Cosmos \& Physics
  Department, Penn State, University Park, PA 16802, U.S.A.}

\begin{abstract}

The very early universe provides the best arena we currently
have to test quantum gravity theories. The success of the
inflationary paradigm in accounting for the observed
inhomogeneities in the cosmic microwave background already
illustrates this point to a certain extent because the paradigm
is based on quantum field theory on the curved cosmological
space-times. However, this analysis excludes the Planck era
because the background space-time satisfies Einstein's
equations all the way back to the big bang singularity. Using
techniques from loop quantum gravity, the paradigm has now been
extended to a self-consistent theory from the Planck regime to
the onset of inflation, covering some 11 orders of magnitude in
curvature. In addition, for a narrow window of initial
conditions, there are departures from the standard paradigm,
with novel effects, such as a modification of the consistency
relation involving the scalar and tensor power spectra and a
new source for non-Gaussianities. Thus, the genesis of the
large scale structure of the universe can be traced back to
quantum gravity fluctuations \emph{in the Planck regime}. This
report provides a bird's eye view of these developments for the
general relativity community.

\end{abstract}


\maketitle

\section{Introduction}
\label{s1}

In this conference, Professor Bicak and others described the
ideas that Einstein developed in Prague during 1911-12. From
then until 1915 he worked largely by himself on the grand
problem of extending the reach of special relativity to
encompass gravity. Finally, in November 1915, he provided us
with the finished theory. For almost a century, the relativity
community has been engaged in understanding the astonishingly
rich physics it contains, testing it ever more accurately, and
applying it to greater and greater domains of astrophysics and
cosmology. The theory has so many marvelous features.
Amazingly, the field equations turned out to provide an
elliptic-hyperbolic system with a well-posed initial value
problem. After many decades, we realized that the total mass of
an isolated system is a well defined geometric invariant and,
furthermore, positive if the local energy density of matter is
positive. The theory naturally admits cosmological solutions in
which the universe is expanding, just as the observations tell
us. It admits black hole solutions that model the engines for
the most energetic phenomena seen in the universe. None of
these fascinating features that we now regard as fundamental
consequences were part of Einstein's motivation during his
quest which he described as \emph{``one of the most exciting
and exacting times of my life"} \cite{ae}. He essentially
handed to us the finished product on a platter. We have been
engaged in uncovering the numerous hidden treasures it contains
by working out the philosophical, mathematical, physical,
astronomical and cosmological consequences of the new paradigm.

But we know that the theory is incomplete. Indeed, it exhibits its
own fundamental limitations through singularities where space-time
ends and general relativistic physics comes to a halt. We also
understand that this occurs because general relativity ignores
quantum physics. Perhaps the most outstanding example is the
prediction of the big bang. If we go back in time, \emph{much}
before we reach the singularity, matter densities exceed the nuclear
density, $\sim 10^{14} - 10^{15} {\rm gms/cc}$, where we definitely
know that quantum properties of matter dominate. Since gravity
couples to matter, the conceptual paradigm of general relativity
becomes inadequate. If we go further back in time, general
relativity presents us with an epoch in which densities reach $\sim
10^{94} {\rm gms/cc}$. This is the Planck scale and now physics of
general relativity becomes inadequate not only conceptually but also
in practice. In this regime we expect gross departures from
Einstein's theory. Just as it is totally inadequate to use Newtonian
mechanics to explore physics near the horizon of a solar mass black
hole, it is incorrect to trust general relativity once the matter
density and space-time curvature enter the Planck regime. Thus,
\emph{big bang is a prediction of general relativity in a domain in
which it is simply invalid}. Normally physicists do not advertise
such predictions of theories. But unfortunately they often seem to
make an exception for the big bang. One hears statements like `the
cosmic microwave background (CMB) is a fingerprint of the big bang'.
But in the standard scenario, CMB refers to a time some 380,000
years after the putative big bang. Existence or even the detailed
features of CMB have no bearing on whether the big bang with
\emph{infinite} matter density and curvature ever occurred. Indeed,
as we will see, loop quantum cosmology (LQC) has no big bang
singularity and yet reproduces these features. What about inflation?
In the standard scenario, it is supposed to have commenced `only'
$10^7$ Planck seconds after the big bang. Does its success not imply
that there was a big bang? It does \emph{not} because the matter
density and curvature at the onset of inflation are only
$10^{-11}-10^{-12}$ times the Planck scale. Indeed, this is why one
can use Einstein's equations and quantum field theory (QFT) on
Friedmann, Lema\^itre, Robertson, Walker (FLRW) solutions in the
analysis of inflation. Inflationary physics by itself cannot say
what really happened in the Planck regime and, again, as we will
see, is compatible with the LQC prediction that there was no big
bang singularity.

Thus, to know what really happened in the Planck regime and go
beyond the singularities predicted by general relativity, we need a
viable quantum theory of gravity. Since the search for this theory
has been ongoing for decades, justifiably, there is sometimes a
sentiment of pessimism in the general relativity circles. In my
view, this is largely because one judges progress using the
criterion of general relativity. In a masterful stroke, Einstein
gave us the final theory and we have been happily engaged in
investigating its content. It seems disappointing that this has not
happened with quantum gravity. But progress of physical theories has
more often mimicked the development of quantum theory rather than
general relativity. More than a century has passed since Planck's
discovery that launched the quantum. Yet, the theory is incomplete.
We do not have a satisfactory grasp of the foundational issues,
often called the `measurement problem', nor do we have a single
example of an interacting QFT in 4 dimensions. A far cry from what
Einstein offered us in 1915! Yet, no one would deny that quantum
theory has been extremely successful; indeed, much more so than
general relativity.

Thus, while it is tempting to wait for another masterful stroke like
Einstein's to deliver us a finished quantum gravity theory, it is
more appropriate to draw lessons from quantum theory. There,
progress occurred by focussing not on the `final, finished' theory,
but on concrete physical problems where quantum effects were
important. It would be more fruitful to follow this path in quantum
gravity. Indeed, even though we are far from a complete theory,
advances can occur by focusing on specific physical problems and
challenges.

Over the last several years, research in loop quantum gravity (LQG)
has been driven by this general philosophy. In addition to seeking a
completion of the general program based on connection variables,
spin networks and spin foams, more and more effort is now focused on
specific physical problems where quantum gravity effects are
expected to be important. The idea behind this research is to
\emph{first truncate general relativity (with matter) to sectors
tailored to specific physical problems, and then pass to quantum
theory using the background independent methods based on the
specific quantum geometry that underlies LQG.} This strategy of
focusing on specific problems of quantum gravity also distinguishes
LQG from string theory in terms of their main trust in the last few
years. In string theory, the focus has shifted to \emph{using} the
well-understood parts of gravity to explore other areas of physics
---use of the AdS/CFT hypothesis to understand the strong coupling
regime of QCD, to gain insights into hydrodynamics and tackle
the strong coupling problems in mathematical physics to better
understand condensed matter systems such as high temperature
super-conductivity. The LQG community, on the other hand, has
continued to tackle the long standing problems of quantum
gravity per se ---absence of a space-time in the background,
the problem of time, fate of cosmological singularities in the
quantum theory, quantum geometry of horizons, and derivation of
the graviton propagator in a background independent setting.

The goal of my talk was to report the advances in the cosmology
of the very early universe that have resulted from a continued
application of the truncation strategy in LQG. Of course, both
the talk and this report can only provide a bird's eye view of
these developments. The results I reported are based largely on
joint work with Alejandro Corichi, Tomasz Pawlowski and
Parampreet Singh \cite{aps1,aps2,aps3,acs,pa,aps4} on the
singularity resolution in cosmology;  with David Sloan
\cite{as2,as3} on effective LQC dynamics tailored to inflation,
with Wojciech Kaminski and Jerzy Lewandowski \cite{akl} on QFT
on quantum space-times; and especially with Ivan Agullo and
William Nelson on extension of the cosmological perturbation
theory to the Planck regime and its application to inflation
\cite{aan1,aan2,aan3}. (For a short overview of the last three
papers, see \cite{param}.) Therefore, there is a large overlap
with the material covered in these original references.
Finally, by now there are well over a 1000 papers on LQC which
include several investigations of inflationary dynamics. What I
can cover constitutes only a very small fraction of what is
known. For reviews on results until about a year ago, see, e.g.
\cite{mbrev,asrev}.

\section{Setting the Stage}
\label{s2}

Perhaps the most significant reason behind the rapid and spectacular
success of quantum theory, especially in its early stage, is the
fact that there was already a significant accumulation of relevant
experimental data, and further experiments to weed out ideas could
be performed on an ongoing basis. Unfortunately this is \emph{not}
the case for quantum gravity simply because theory has raced far
ahead of technology. Indeed, even in the classical regime, we still
lack detailed tests of general relativity in the strong field
regime!

Currently, the early universe offers by far the best arena to test
various ideas on quantum gravity. Most scenarios assume that the
early universe is well described by a FLRW solution to Einstein's
equations with suitable matter, \emph{together with} first order
perturbations. The background is treated classically, as in general
relativity, and the perturbations are described by \emph{quantum
fields}. Thus, the main theoretical ingredient in the analysis are:
cosmological perturbation theory and QFT on FLRW space-times. It is
fair to say that among the current scenarios, the inflationary
paradigm has emerged as the leading candidate. In addition to the
common assumption described above, this scenario posits:

\begin{itemize}

\item Sometime in its early history, the universe underwent a
    phase of rapid expansion. This was driven by the slow roll
    of a scalar field in a suitable potential causing the Hubble
    parameter to be nearly constant.

\item Fourier modes of the quantum fields representing
    perturbations were initially in a specific state, called the
    Bunch-Davies (BD) vacuum, for a certain set of co-moving
    wave numbers $(k_o, 2000k_o)$ where the physical wave length
    of the mode $k_o$ equals the radius $R_{\rm LS}$ of the
    observable universe at the surface of last scattering.%
\footnote{Strictly speaking, the BD vacuum refers to de
Sitter space; it is the unique `regular' state which is
invariant under the full de Sitter isometry group. During
slow roll, the background FLRW geometry is only
approximately de Sitter whence there is some ambiguity in
what one means by the BD vacuum. One typically assumes that
all the relevant modes are in the BD state (tailored to) a
few e-foldings before the mode $k_o$ leaves the Hubble
horizon. Throughout this report, by BD vacuum I mean this
state.}

\item Soon after any mode exits the Hubble radius, its quantum
    fluctuation can be regarded as a classical perturbation and
    evolved via linearized Einstein's equations.
\end{itemize}

One then evolves the perturbations from the onset of the slow
roll till the end of inflation using QFT on FLRW space-times
and calculates the power spectrum (see, e.g.,
\cite{ll-book,sd-book,vm-book,sw-book,gr-book}). When combined
with standard techniques from astrophysics to further evolve
the results to the surface of last scattering, one finds that
they are in excellent agreement with the inhomogeneities seen
in the CMB. Supercomputer simulations have shown that these
inhomogeneities serve as seeds for the large scale structure in
the universe. Thus, in a precise sense, the origin of the
qualitative features of the observed large scale structure can
be traced back to the fluctuations in the quantum vacuum at the
onset of inflation. This is both intriguing and very
impressive.

Over the years, the inflationary paradigm has witnessed
criticisms from the relativity community, most eloquently
expressed by Roger Penrose (see, e.g., \cite{rp1}). However,
these criticisms refer to the motivations that were originally
used by the proponents, rather than to the methodology
underlying its success in accounting for the CMB
inhomogeneities. There are plenty of examples in fundamental
physics where the original motivations turned out not to be
justifiable but the idea was highly successful. I share the
view that the basic assumptions, listed above, are neither
`obvious' nor have they been justified from first principles.
However, the success of the inflationary paradigm with CMB
measurements is nonetheless impressive because one `gets much
more out than what one puts in'.

In spite of this success, however, the inflationary scenario is
conceptually incomplete in several respects. (For a cosmology
perspective on these limitations see e.g. \cite{brandenberger}.) In
particular, as Borde, Guth and Vilenkin \cite{bgv} showed,
inflationary space-times inherit the big-bang singularity in spite
of the fact that the inflaton violates the standard energy
conditions used in the original singularity theorems \cite{hebook}.
As we discussed in section \ref{s1}, this occurs because one
continues to use general relativity even in the Planck regime in
which it is simply not applicable. One expects new physics to play a
dominant role in this regime, thereby resolving the singularity and
significantly changing the very early history of the universe. One
is therefore led to ask: Will inflation arise naturally in the
resulting deeper theory? Or, more modestly, can one at least obtain
a consistent quantum gravity extension of this scenario?

The open-ended nature of the inflationary paradigm has three facets.
First, there are issues whose origin lies in particle physics. Where
does the inflaton come from? How does potential arise? Is there a
single inflaton or many? If many, what are the interactions between
them? Since the required mass of the inflaton is very high, above
$10^{12} {\rm Gev}$, the fact that we have not seen it at CERN does
not mean it cannot exist. But in the inflationary scenario this is
the only matter field in the early universe and particles of the
standard model are supposed to be created during `reheating' at the
end of inflation when the inflaton is expected to roll back and
forth around its minimum. However, how this happens is not at all
well-understood. What are the admissible interactions between the
inflaton and the standard model particles which causes this decay?
Does the decay produce the correct abundance of the standard model
particles? These questions with origin in particle physics are wide
open.

The second issue is the quantum to classical transition referred to
in the last assumption of standard inflation. In practice one
calculates the expectation values of perturbations and the two point
function at the end of inflation and assumes one can replace the
actual quantum state of perturbations with a Gaussian statistical
distribution of classical perturbations with the mean and variance
given by the quantum expectation value and the 2-point function. As
a calculational devise this strategy works very well. However, what
happens physically? While this issue has drawn attention, we do not
yet have a clear consensus on the actual, detailed physics that is
being approximated in the last assumption.

The third set of issues have their origin in quantum gravity.
In the standard inflationary scenario, one specifies initial
conditions at the onset of inflation and then evolves the
quantum perturbations. As a practical strategy, something like
this is unavoidable within general relativity. Ideally one
would like to specify the initial conditions at `the
beginning', but one simply cannot do this because the big bang
is singular. Furthermore, since the curvature at the onset of
inflation is some $10^{-11} - 10^{-12}$ times the Planck scale,
by starting calculations there, one bypasses the issue of the
correct Planck scale physics. But this is just an astute
stopgap measure. Given any candidate quantum gravity theory,
one can and \emph{has to} ask whether one can do better. Can
one meaningfully specify initial conditions in the Planck
regime? In a viable quantum gravity theory, this should be
possible because there would be no singularity and the Planck
scale physics would be well-controlled. If so, in the
systematic evolution from there, does a slow roll phase
compatible with the 7 year WMAP data \cite{wmap} arise
\emph{generically} or is an enormous fine tuning needed? One
could argue that it is acceptable to use fining tuning because,
after all, the initial state is very spatial. If so, can one
provide physical principles that select this special state? In
the standard inflationary scenario, if we evolve the modes of
interest back in time, they become trans-Planckian. Is there a
QFT on \emph{quantum} cosmological space-times needed to
adequately handle physics at that stage? Can one \emph{arrive
at} the BD vacuum (at the onset of the WMAP slow roll) staring
from natural initial conditions at the Planck scale?

In this report, I will not address the first two sets of issues.
Rather, the focus will be on the incompleteness related to the third
set, i.e., on quantum gravity. Systematic advances within LQC over
the past six years have provided a viable extension of the
inflationary scenario all the way to the Planck regime. This
extension enables us to answer in detail most of the specific
questions posed above. To arrive at a coherent extension, LQC had to
develop a conceptual framework, mathematical tools and high
precision numerical simulations because the issues are so diverse:
The meaning of time in the Planck regime; the nature of quantum
geometry in the cosmological context; QFT on \emph{quantum}
cosmological space-times; renormalization and regularization of
composite operators needed to compute stress energy and back
reaction; and, relation between theory and the WMAP data.

A consistent theoretical framework to deal with cosmological
perturbations on quantum FLRW space-times now exists \cite{aan2}.
Starting with `natural' initial conditions in the Planck regime, one
can evolve the quantum perturbations on quantum FLRW backgrounds and
study in detail the pre-inflationary dynamics \cite{aan1,aan3}.
Detailed numerical simulations have shown that the \emph{predictions
are in agreement with the power spectrum and the spectral index
reported in the 7 year WMAP data.} However, there is also a small
window in the parameter space where the initial state at the onset
of inflation differs sufficiently from the BD vacuum assumed in
standard inflation to give rise to new effects. 
These are the prototype observable signatures of pre-inflationary
dynamics. In this sense, \emph{LQC offers the possibility of
extending the reach of cosmological observations to the deep Planck
regime of the early universe.}

\section{Why Pre-inflationary Dynamics Matters}
\label{s3}

\begin{figure}[]
 \begin{center}
  \includegraphics[scale=0.65]{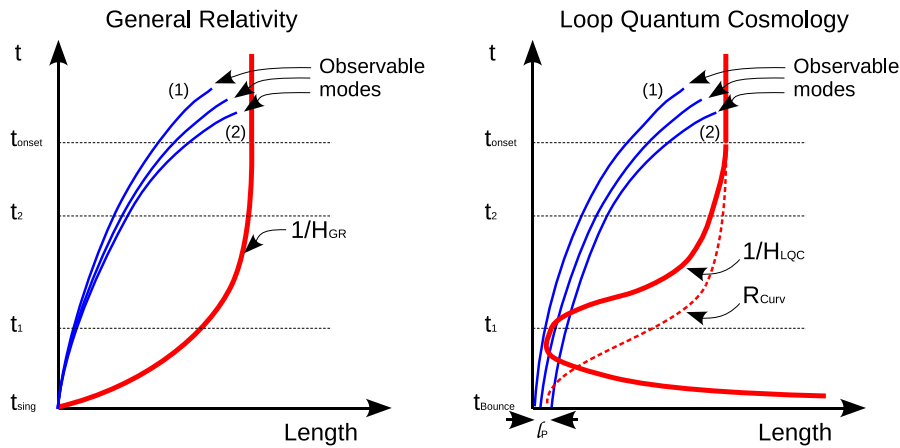}
  \caption{\label{fig1} Schematic time evolution of the Hubble radius
  (red solid line on the right in each panel) and of wave lengths of three
  modes seen in the CMB (three solid blue lines in each panel).
  Credits: W. Nelson.\,\,
  \emph{Left Panel: General relativity.} The modes of interest have wave lengths
  less than the Hubble radius $1/H_{\rm GR}$ all the way from the big bang
  ($t_{\rm sing}$) until after the onset of slow roll.\\
  \emph{Right Panel: LQC.} The Hubble radius diverges at the big bounce
  ($t_{\rm Boun}$), decreases rapidly to reach its minimum in the
  deep Planck era and then increases monotonically. Because of this, modes seen
  in the CMB can have wave lengths larger than the Hubble radius $1/H_{\rm LQC}$
  in the very early universe. Detailed analysis shows that what really matters
  is the curvature radius $R_{\rm curv}$ shown schematically by the dashed red
  line rather than the Hubble radius $1/H_{\rm LQC}$. But again the modes can exit
  the curvature radius in the Planck regime and, if they do, they are excited during
  the pre-inflationary evolution. They will not be in the BD vacuum at the onset of
  slow roll inflation.}
\end{center}
\end{figure}

It is often claimed that pre-inflationary dynamics will not change
the observable predictions of the standard inflationary scenario.
Indeed, this belief is invoked to justify why one starts the
analysis just before the onset of the slow roll. The belief stems
from the following argument, sketched in the left panel of
Fig.~\ref{fig1}. If one evolves the modes that are seen in the CMB
\emph{back} in time starting from the onset of slow roll, their
physical wave lengths $\lambda_{\rm phy}$ continue to remain within
the Hubble radius $1/H_{\rm GR}$ all the way to the big bang.
Therefore, one argues, they would not experience curvature and their
dynamics would be trivial all the way from the big bang to the onset
of inflation; because they are not `exited', all these modes would
be in the BD vacuum at the onset of inflation. However, this
argument is flawed on two accounts. First, if one examines the
equation governing the evolution of these modes, one finds that what
matters is the curvature radius ${R}_{\rm curv} =
\sqrt{6/\mathfrak{R}}$ determined by the Ricci scalar
$\mathfrak{R}$, and not the Hubble radius. The two scales are
equivalent only during slow roll on which much of the intuition in
inflation is based. However, in general they are quite different
from one another. Thus we should compare $\lambda_{\rm phy}$ with
${R}_{\rm curv}$ in the pre-inflationary epoch. The second and more
important point is that the pre-inflationary evolution should not be
computed using general relativity, as is done in the argument given
above. One has to use an appropriate quantum gravity theory since
the two evolutions are expected to be \emph{very} different in the
Planck epoch. Then modes that are seen in the CMB could well have
$\lambda_{\rm phy} \gtrsim {R}_{\rm curv}$ in the pre-inflationary
phase. If this happens, these modes \emph{would be} excited and the
quantum state at the onset of the slow roll could be quite different
from the BD vacuum. Indeed, the difference could well be so large
that the amplitude of the power spectrum and the spectral index are
incompatible with WMAP observations. In this case, that particular
quantum gravity scenario would be ruled out. On the other hand, the
differences could be more subtle: the new power spectrum for scalar
modes could be compatible with observations  but there may be
departures from the standard predictions that involve tensor modes
or higher order correlation functions of scalar modes, changing the
standard conclusions on non-Gaussianities
\cite{holman-tolley,agullo-parker,ganc,agullo-navarro-salas-parker}.
In this case, the quantum gravity theory would have interesting
predictions for future observational missions. Thus,
pre-inflationary dynamics can provide an avenue to confront quantum
gravity theories with observations.

These are not just abstract possibilities. The right panel of
Fig.~\ref{fig1} shows schematically the situation in LQC. (For the
precise behavior obtained from numerical simulations, see Fig. 1 in
\cite{aan3}.) The wave lengths of some of the observable modes
\emph{can} exit the curvature radius during pre-inflationary
dynamics, whence there are departures from the standard predictions
(which turn out to be of the second type in the discussion above).

So far we have focused only on why a common argument suggesting that
pre-inflationary dynamics cannot have observational consequences is
fallacious. At a deeper level, pre-inflationary dynamics matters
because of a much more general reason: It is important to know if
inflationary paradigm is part of a conceptually coherent framework
encompassing the quantum gravity regime. Can one trust the standard
scenario in spite of the fact that the modes it focuses on become
trans-Planckian in the pre-inflationary epoch? Does one have to
artificially fine-tune initial conditions in the Planck regime to
arrive at the BD vacuum? Do initial conditions for the background in
the Planck regime naturally give rise to solutions that encounter
the desired inflationary phase some time in the future evolution? To
investigate any one of these issues, one needs a reliable theory for
pre-inflationary dynamics and also good control on its predictions.

\section{The LQG Strategy}
\label{s4}

LQG offers an attractive framework to investigate
pre-inflationary dynamics because its underlying quantum
geometry becomes important at the Planck scale and leads to the
resolution of singularities in a variety of cosmological
models. In particular the following cosmologies have been
investigated in detail:  the k=0 and k=1 FLRW models are
discussed in \cite{mb1,abl,aps1,aps2,aps3,acs,ps,apsv,warsaw1};
a non-zero cosmological constant is included in
\cite{bp,kp1,pa}; anisotropic are discussed via Bianchi I, II
and IX models in \cite{awe2,madrid-bianchi,awe3,we}; and the
inhomogeneous Gowdy models ---that have attracted a great deal
of attention in mathematical general relativity--- were studied
in \cite{hybrid1,hybrid2,hybrid3,hybrid4,hybrid5}. In all
cases, the big bang singularity is resolved and replaced by
quantum bounces. It is therefore natural to use LQC as the
point of departure for extending the cosmological perturbation
theory.

In the standard perturbation theory, one begins with linearized
solutions of Einstein's equations on a FLRW background.
Unfortunately, we cannot mimic this procedure because in LQG we
do not yet have the analog of full Einstein's equations that
one could perturb. But one can adopt the \emph{truncation
strategy} discussed in section \ref{s1}. Thus, one starts with
a truncation $\ps_{\tr}$ of the phase space $\ps$ of general
relativity, tailored to the linear perturbations off FLRW
backgrounds. Furthermore since we are interested in the issue
of whether the inflationary framework admits a quantum gravity
extension, the matter source will be just a scalar field $\phi$
with the simplest, i.e. quadratic, potential $V(\phi) = (1/2)
m^2\phi^2$. Thus, $\ps_{\tr}$ is given by $\ps_{\tr} =
\ps_o\times \ps_1$ where $\ps_o$ is the 4-dimensional FLRW
phase space, with the scale factor $a$ and the homogeneous
inflaton $\phi$ as configuration variables, and $\ps_1$ is the
phase space of gauge invariant first order perturbations
consisting of a scalar mode and two tensor modes. Since the
background fields are homogeneous, it is simplest to assume
that the perturbations are purely inhomogeneous. Thus, regarded
as a sub-manifold of the full phase space $\ps$,\, $\ps_{\tr}$
is the normal bundle over $\ps_o$.

As usual, for perturbations one can freely pass between real space
and momentum space using Fourier transforms of fields in co-moving
coordinates. For pre-inflationary dynamics, we work with the
Mukhanov-Sasaki variables, denoted by  $\Q_{\vk}$, because they are
well-defined all the way from the bounce to the onset of slow roll.%
\footnote{The curvature perturbations $\R_{\vk}$ fail to be
well-defined at the `turning point' where $\dot\phi =0$, which
occurs during pre-inflationary dynamics. However, they are much more
convenient for relating the spectrum of perturbations at the end of
inflation with the CMB temperature fluctuations. Therefore, we first
calculate the power spectrum $\mathcal{P}_{\Q}$ for Mukhanov-Sasaki
variable $\Q_{\vk}$ and then convert it to $\mathcal{P}_{\R}$,
reported in Fig.~\ref{fig3}.}
We denote the two tensor modes collectively by $\T_{\vk}$. This
structure is the same as that used in standard inflation
\cite{langlois}.

New features appear in the next step: In the passage to quantum
theory, we work with the \emph{combined system}, i.e., with all of
$\ps_{\tr}$. Therefore, we are naturally led a theory in which not
only the perturbations but even the background geometry is quantum.
Rather than having quantum fields $\h{\Q}$ and $\h\T$ propagating on
a classical FLRW space-time, they now propagate on a \emph{quantum}
FLRW geometry.

Thus, the strategy to truncate the classical phase space and
then pass to quantum theory using LQG techniques leads to a
novel quantum theory. The total Hilbert space is a tensor
product, $\H = \H_o\otimes\H_1$, where $\H_o$ is the space of
wave functions $\Psi_o$ describing a quantum FLRW geometry and
$\H_1$ is the space of quantum states $\psi$ of perturbations.
The first task is to construct the Hilbert space $\H_o$ of
physical states $\Psi_o(a,\phi)$, by imposing the Hamiltonian
constraint on the quantum theory of the homogeneous sector
$\ps_o$. The second task is to study quantum dynamics of fields
$\h{\Q}$ and $\h{\T}$ on the \emph{quantum} geometry
encapsulated in $\Psi_o(a,\phi)$. In particular we have to
introduce the Hilbert space $\H_1$ of wave functions
$\psi(\Q_{\vk}, \T_{\vk})$ of perturbations and develop
techniques to calculate the 2-point functions on $\H_1$ that
are needed to obtain the scalar and the tensor power spectra.
The final task is to check the self-consistency of the
truncation strategy with which we began. Already in the
classical theory, the truncated phase space $\ps_{\tr}$ is
useful only so long as the back reaction can be neglected.
Therefore, in the quantum theory, we have to check that $\H$
admits solutions $\Psi_o\otimes\psi$ in which the energy
density of perturbations is negligible compared to that in the
background all the way from the LQC bounce to the onset of slow
roll. On the analytical side, this requires the introduction of
suitable regularization and renormalization techniques for
quantum fields $\h{\Q}$ and $\h{\T}$ propagating on the quantum
background $\Psi_o$. On the numerical side, one has to devise
accurate numerical methods to calculate the energy density in
perturbations with sufficient precision during the evolution
all the way from the bounce to the onset of inflation, as the
background energy density falls by some 11 orders of magnitude.

These tasks have been carried out in \cite{aan1,aan2,aan3} using
earlier results obtained in \cite{aps3,acs,as2,as3,akl,pa,aps4}. The
next two sections provide a flavor of this analysis.

\section{Analytical Aspects}
\label{s5}

\bu \emph{Background Quantum Geometry:} In the classical theory,
dynamics on $\Gamma_o$ is generated by the single, homogeneous,
Hamiltonian constraint, $\mathbb{C}_o =0$. Each dynamical trajectory
on $\ps_o$ represents a classical FLRW space-time. In quantum
theory, physical states are represented by wave functions
$\Psi_o(a,\phi)$ satisfying the quantum constraint
$\h{\mathbb{C}}_o\, \Psi_o =0$. Each of these solutions represents a
quantum FLRW geometry.

\begin{figure}
\begin{center}
\includegraphics[width=4.7in,height=2.7in]{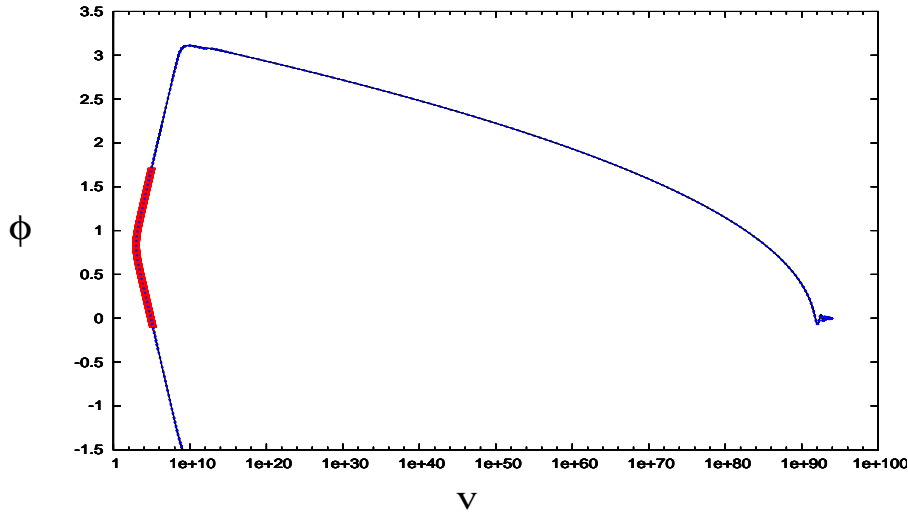}
\caption{\label{fig2} An effective LQC trajectory in presence of an
inflation with a quadratic potential $(1/2) m^2\phi^2$, where
the value $m = 6.1 \times 10^{-6} m_{\rm Pl}$ of the mass is calculated
from the 7 year WMAP data (source \cite{aps4}). Here $V \sim a^3$ is the
volume of a fixed fiducial region. The long (blue) sloping line at
the top depicts slow roll inflation. As $V$ decreases (right to left),
we go back in time and the inflaton $\phi$ first climbs up the potential,
then turns around and starts going descending. In classical general relativity,
volume would continue to decrease until it becomes zero, signalling the big
bang singularity. In LQC, the trajectory bounces at $\phi \sim 0.95$ and volume
never reaches zero; the entire evolution is non-singular.}
\end{center}
\end{figure}

We are interested in those solutions $\Psi_o$ which remain
sharply peaked on classical FLRW solutions at late times. In
the sector of the theory that turns out to be physically most
interesting \cite{aan3}, these states remain sharply peaked all
the way up to the bounce but in the Planck regime they follow
certain effective trajectories which include quantum
corrections \cite{asrev,aps4}. In particular, rather than
converging on the big bang singularity, as classical FLRW
solutions do, they exhibit a bounce when the density reaches
$\rcr \approx 0.41 \rho_{\rm Pl}$ (see Fig.~\ref{fig2}). It
turns out that each (physically distinct) effective solution is
completely characterized by the value $\phi_{\B}$ that the
inflaton assumes at the bounce. \emph{This value turns out to
be the key free parameter of the theory.} Finally, we need full
quantum evolution from the bounce only until the density and
curvature fall by a factor of, say, $10^{-3}-10^{-4}$. After
that, the background can be taken to follow the general
relativity trajectory to a truly
excellent approximation.%
\footnote{During this phase, the scalar field is monotonic in time
in the effective trajectory. Therefore we can use the scalar field
as an `internal' or `relational' time variable with respect to which
the background scale factor (and curvature) as well as perturbations
evolve. This interpretation is not essential but very helpful in
practice because of the form of the Hamiltonian constraint
$\h{\mathbb{C}}_o\, \Psi_o =0$ (for details, see e.g.
\cite{asrev}).}
(For details, see \cite{aps3,acs,aps4}).

\bu \emph{Dynamics of Perturbations:} There is an important subtlety
which is often overlooked in the quantum gravity literature:
Dynamics of perturbations is \emph{not} generated by a constraint,
or, indeed by \emph{any} Hamiltonian. On the truncated phase space
$\ps_{\tr}$, the dynamical trajectories are tangential to a vector
field $X^\alpha$ of the form $X^\alpha = \Omega^{\alpha\beta}_o
\partial_\beta \mathbb{C}_o + \Omega^{\alpha\beta}_1\partial_\beta
\mathbb{C}_2^\prime$ where $\Omega_o$ and $\Omega_1$ are the
symplectic structures on $\Gamma_o$ and $\Gamma_1$, and
$\mathbb{C}_2^\prime$ is the part of the second order
Hamiltonian constraint function in which only terms that are
quadratic in the first order perturbations are kept (ignoring
terms which are linear in the second order perturbations).
$X^\alpha$ fails to be Hamiltonian on $\Gamma_{\tr}$ because
$\mathbb{C}_2^\prime$ depends not only on perturbations but
also background quantities. However, given a dynamical
trajectory $\gamma_o(t)$ on $\Gamma_o$ and a perturbation at a
point thereon, $X^\alpha$ provides a canonical lift of
$\gamma_o(t)$ to the total space $\Gamma_{\tr}$, describing the
evolution of that perturbation along $\gamma_o(t)$. In
space-time language this corresponds to first fixing a
background FLRW solution and then solving for the (first order)
perturbations propagating on that background.

Therefore, in the quantum theory, dynamics of the combined system
\emph{cannot be obtained by simply imposing a quantum constraint} on
the wave functions $\Psi_o\otimes\psi$ of the combined system. One
has to follow a procedure similar to what is done in the classical
theory. Thus, one first obtains a background quantum geometry
$\Psi_o$ by solving $\h{\mathbb{C}}_o\, \Psi_o (a,\phi) =0$,
specifies the quantum state $\psi(\Q_{\vk}, \T_{\vk})$ of the
perturbation at, say, the bounce time, and evolves it using the
operator $\h{\mathbb{C}}_2^\prime$. The resulting state $\Psi(a,
\Q_{\vk}, \T_{\vk}, \phi)$ describes the evolution of the quantum
perturbations $\psi$ on the quantum geometry $\Psi_o$ in the
Schr\"odiger picture. (For details, see \cite{aan2}).

\bu \emph{Trans-Planckian Issues:} Quantum perturbations $\h{\Q},\,
\h{\T}$ propagate on quantum geometries $\Psi_o$ which are all
regular, free of singularities. Thus, the \emph{the framework is
tailored to cover the Planck regime.} What is the status of the
`trans-Planckian problems' which are associated with modes of
trans-Planckian frequencies in heuristic discussions? To probe this
issue one has to first note that the quantum Riemannian geometry
underlying LQG is quite subtle \cite{alrev,crbook,ttbook}: in
particular, while there is a minimum non-zero eigenvalue of the area
operators, the area gap, there is no volume gap, even though their
eigenvalues are also discrete
\cite{rs,al5}.%
\footnote{Properties of the eigenvalues of length operators
\cite{tt-length,eb-length,msy-length} have not been analyzed in
comparable detail. But since their definitions involve volume
operators, it is expected that there would be no `length gap'.}
As a consequence, there is no fundamental obstacle preventing the
existence trans-Planckian modes of perturbations in our truncated
theory. Indeed, in the homogeneous LQC models that have been
analyzed in detail, the momentum $p_{(\phi)}$ of the scalar field
$\phi$ is generally \emph{huge} in Planck units. This poses no
problem and, in particular, on the physical Hilbert space the total
energy density is still guaranteed to be bounded by $\rcr$ (see,
e.g. \cite{asrev}). Similarly, perturbations $\h{\Q},\, \h{\T}$ of
our truncated theory are permitted to acquire trans-Planckian
momenta. The real danger is rather that, in presence of such modes,
the \emph{energy density} in perturbations may fail to be negligible
compared to that in the quantum background geometry. This issue is
extremely non-trivial, especially in the Planck regime. If the
energy density does become comparable to that in the background,
then we would not be able to neglect the
back-reaction and our truncation would fail to be self-consistent.%
\footnote{Of course, this would not imply that the inflationary
scenario does not admit an extension to the Planck regime. But to
obtain it one would then have to await the completion of a
\emph{full} quantum gravity theory.}
\emph{This is the trans-Planckian problem we face in our theory of
quantum perturbations on inflationary quantum geometries.} To
address it we need regularization and renormalization methods to
compute energy density for quantum fields on quantum FLRW
geometries. (For details, see \cite{aan2,aan3}).

\bu \emph{An Unforeseen Simplification:} As we just noted, the
underlying FLRW quantum geometry provides the necessary control
on calculations in the deep Planck regime. However, it
confronts us with a new challenge of developing the
mathematical theory of quantum fields on \emph{quantum}
geometries. At first this problem seems formidable. But
fortunately there is a key simplification within the test field
approximation we are using in the truncated theory
\cite{akl,aan2}: Mathematically the evolution of $\h{\Q},
\h{\T}$ on any one of our quantum geometries $\Psi_o$ is
completely equivalent to that of these fields propagating on a
dressed, effective metric $\t{g}_{ab}$ constructed from $\Psi_o$.%
\footnote{For scalar modes, the classical equation of motion
involves also `an external potential' $\g$. This has also to be
replaced by a dressed effective potential $\t{\g}$. for details, see
\cite{aan3}.}
Note that $\t{g}_{ab}$ contains quantum corrections and does not
satisfy Einstein's equation. Indeed, it does not even satisfy the
effective equations of LQC because, whereas the effective
trajectories follow the `peak of $\Psi_o$', $\t{g}_{ab}$ also knows
about certain fluctuations encoded in $\Psi_o$.%
\footnote{While this difference is conceptually important, because
the states $\Psi_o$ of interest are so sharply peaked, in practice
the deviations from effective trajectories are small even in the
Planck regime. Of course the deviations from classical solutions are
enormous in the Planck regime because $\t{g}_{ab}$ is non-singular.}
Nonetheless, since $\t{g}_{ab}$ is a smooth metric with FLRW
symmetries, it is now possible to use the rich machinery of QFT on
cosmological space-times to analyze the dynamics of $\h{\Q}, \h{\T}$
in detail. In addition, one can now make use of the powerful
technique of adiabatic regularization that has been developed over
some three decades
\cite{parker66,pf,birrell,anderson-parker,sf-book,parker-book}. In
particular, by restricting ourselves to states $\psi$ of
perturbations which are of 4th adiabatic order, one can compute the
expectation values of energy density. This provides a clear avenue
to face the true trans-Planckian problem, i.e., to systematically
test if the truncation approximation is valid.

This remarkable simplification occurs because the dynamics of test
quantum fields is not sensitive to all the details of the
probability amplitude for various FLRW metrics encapsulated in
$\Psi_o$; it experiences only to a few moments of this distribution.
The phenomenon is analogous to the propagation of light in a medium
where all the complicated interactions of the Maxwell field with the
atoms in the medium can be captured just in a few parameters such as
the refractive index. (For details, see \cite{akl,aan2,aan3}).

\bu \emph{Initial Conditions:} In the Schr\"odinger picture,
the above simplification enables us to evolve the quantum state
$\psi$ of perturbations. But we still have to specify the
initial conditions. Since the big bang of general relativity is
replaced by the big bounce in LQC, it is natural to specify
them at the bounce. Now, in the truncation approximation,
perturbation is treated as a test field. Therefore, it is
appropriate to assume that the initial state has the form
$\Psi_o \otimes \psi$ at the bounce. Furthermore this simple
tensor product form will be preserved under dynamics so long as
the back reaction due to the perturbation remains negligible.

Let us begin with $\Psi_o$. In the effective theory, phase
space variables are subject to certain constraints at the
bounce. We assume that $\Psi_o$ is sharply peaked at a point on
this constraint surface (with small fluctuations in each of the
two `conjugate' variables). At the bounce, the allowed range of
$\phi$ is finite but large, $|\phi_{\B}| \in (0,\, 7.47 \times
10^{5})$ in Planck units. For simplicity, let me consider only
$\phi_{\B} \ge 0$. A detailed analysis of effective solutions
has shown that unless $\phi_{\B} <0.93$, the effective
trajectory necessarily encounters a slow roll phase compatible
with WMAP sometime in the future \cite{as3}. Thus, the peak of
initial $\Psi_o$ is almost unconstrained. However, the
requirement that $\Psi_o$ be peaked is very strong and makes
the initial state of background geometry very special.

For perturbations, we assume the following three conditions on
$\psi$ at the bounce: i) Symmetry: $\psi$ should be invariant
under the FLRW isometry group, i.e., under spatial translations
and rotations. This condition is natural because these are the
symmetries of the background $\Psi_o$ and hence also of
$\t{g}_{ab}$ it determines; ii) Regularity: $\psi$ should be of
4th adiabatic order so that the Hamiltonian operator has a
well-defined action on it; and, iii) The initial renormalized
energy density $\langle \psi|\, \h{\rho}\,|\psi\rangle_{\rm
ren}$ in the perturbation should be negligible compared to the
energy density $\rcr$ in the background. We have an explicit
example, $|\psi\rangle = |0_{\rm obv}\rangle$, of such a state
called the `obvious vacuum of 4th adiabatic order' which has
several attractive properties \cite{aan3}. Furthermore we also
know that, given a state satisfying these properties, there are
`infinitely many' such states in its neighborhood. Thus, the
existence of the desired states is assured. However, in view of
the large freedom that remains, it would be worthwhile to
develop clear-cut physical criteria to cut down this freedom
significantly. This is an open issue, currently under
investigation. (For details, see \cite{aan1,aan2,aan3}).

Let us summarize the analytical framework. The initial
condition for the quantum state $\Psi_o\otimes\psi$ of the
combined system can be easily specified at the bounce in such a
manner that a slow roll inflation compatible with the 7 year
WMAP data is guaranteed in the background geometry. Thanks to
an unforeseen simplification, we can use techniques from QFT on
cosmological space-times to evolve the perturbations $\h{\Q}$
and $\h{\T}$ on the quantum background geometry $\Psi_o$.
Finally, the initial conditions guarantee that the truncation
approximation does hold at the bounce: $\psi$ can be regarded
as a perturbation whose back reaction on $\Psi_o$ is negligible
initially. Furthermore, states are sufficiently regular to
enable us to calculate the energy density in the background and
in the perturbation at all times. Therefore, one can carry out
the entire evolution numerically, calculate the power spectra
and spectral indices and check if the truncation approximation
continues to hold under evolution all the way from the bounce
to the onset of the slow roll.

As discussed in section \ref{s3}, a priori there are several
possible outcomes. Pre-inflationary dynamics could have such a
strong effect that the power spectra and the spectral indices
that result from these calculations are incompatible with the
WMAP observations. In this case, the LQC extension would be
ruled out by observations. It is also possible that the Planck
scale dynamics is such that the back reaction ceases to be
negligible very soon after the bounce making the truncation
strategy inconsistent. One would then have to await full loop
quantum gravity to discuss the early universe in a coherent
fashion. Finally, even if these possibilities do not occur, we
may find that, for observable modes, the state at the onset of
inflation is sufficiently different from the BD vacuum that
there are departures from the standard inflationary predictions
for future observations.

One needs explicit numerical simulations to find out which of
these various a priori possibilities are realized.

\section{Numerical Aspects, Observations and Self-Consistency}
\label{s6}

In this section, numerical values of all physical quantities will be
given in natural Planck units $c$\!\! = \!\!$\hbar$\!\! =
\!\!$G$\!\! = \!\!$1$ (as opposed to the reduced Planck units used
in the cosmology literature where one sets $8\pi G$\!\! = \!\!$1$).
We will use both the conformal time $\t{\eta}$ and the proper (or
cosmic) time $\t{t}$ determined by the dressed effective metric
$\t{g}_{ab}$ via $d\t{s}^2 := \t{g}_{ab} dx^a dx^b = a^2 (-d\t\eta^2
+ d\x^2) = -d\t{t}^2 + a^2 d\x^2$ (where, as usual, $x^a$ are the
co-moving coordinates). This is because the cosmology literature
generally uses conformal time but comparison with general relativity
can be made more transparent in cosmic time by setting it equal to
zero at the big bang in general relativity and at the big
bounce in LQC.\\

\begin{table}
\begin{center}
\begin{tabular}{| c | c | c | c | c | c | }
\hline
$\phi(t_B)$ &  $k_*$ & $\ln k_*$ &
$\lambda_*(t_B)$ &  $t_{k_*}$ & $\ln[a(t_{k_*})/a(t_B)]$
\\ \hline \hline
0.934 & 0.0016 & -6.4 & 4008 & $1.8 \times 10^{5}$& 5.2 \\ \hline
1 & 0.024 & -3.7 & 261 & $5.2 \times 10^{5}$ & 8.0\\ \hline
1.05 & 0.17 & -1.8 & 37.1 &$ 7.6 \times 10^{5}$& 10\\ \hline
1.1 & 1.2 & 0.2 & 5.1 & $1.0 \times 10^{6}$& 12\\ \hline
1.15 & 9.17 & 2.83 & 0.63 & $1.25 \times 10^{6}$& 13.9\\ \hline
1.2 & 70.7  & 4.2 & 0.09 & $1.48 \times 10^{6}$& 16 \\   \hline
\,\,1.3\,\, & \,\,\, $4.58 \times 10^3 $ \,\,\, & \,\, 8.43\,\, & \,\,\,
$1.36\times 10^{-3} $ \,\,\, &\,\,\,\,\,\,$1.97 \times 10^{6}$
\,\,\,\,\,\, & 20.2 \\   \hline
1.5 & $2.7\times 10^7 $ & 17.1 & $2.3\times 10^{-7} $& $2.9 \times
10^{6}$& 28.9\\   \hline
\end{tabular}
\end{center}
\caption{\label{tab:1} This table from \cite{aan3} shows the value
of the reference co-moving momentum $k_\star$ used in the WMAP data,
the corresponding physical wavelength $\lambda_\star(\t{t}_\B)$ at
the bounce, the time $\t{t}(k_{\star})$ at which the mode $k_\star$
exits the Hubble radius during inflation, and
$\ln[{a}(\t{t}(k_{\star}))/{a}(\t{t}_\B)]$, the number of e-folds of
expansion between the bounce and $\t{t}({k_\star})$. We focus on the
range for $\phi_{\B}$ that is relevant to explore whether
pre-inflationary dynamics can lead to deviations from the BD vacuum
at the onset of the slow roll.}
\end{table}

\bu \emph{WMAP Phenomenology:} The 7 year WMAP data \cite{wmap} uses
a reference mode $k_\star \approx 8.58 k_o$ where, as before, $k_o$
is the co-moving wave number of the mode whose physical wave length
equals the radius of the observable universe at the surface of last
scattering. The WMAP analysis provides us with the amplitude
${\mathcal P}_{\R}(k_{\star})$ of the power spectrum and the
spectral index $n_s(k_{\star})$ which encodes the small deviation
from scale invariance, both for the scalar perturbations. The values
are given by
\be \label{observations} {\mathcal P}_{\R}(k_{\star})=(2.430\pm
0.091) \times 10^{-9}\, \quad {\rm and} \quad
n_s(k_{\star})=0.968\pm 0.012\, . \ee
For the quadratic potential considered here, these observational
data provide the following values of the Hubble parameter $H$ and
the slow roll parameter $\epsilon = - \dot{H}/H^2$:
\be \label{H} H(\t\eta(k_{\star}))\,=\, 7.83 \times 10^{-6}\, \quad
{\rm and} \quad \epsilon(\t\eta({k_*}))\, =\,8\times 10^{-3} \, .
\ee
where $\t\eta(k_\star)$ is the conformal time in our dressed
effective metric $\t{g}_{ab}$ at which the mode $k_\star$ exited the
Hubble radius and the `dot' refers to the derivative w.r.t. $\t{t}$.
Since the physical wave length of the mode $k_o$ is 8.58 times
larger, it must have left the Hubble radius $\sim 2$ e-foldings
before $\t\eta(k_\star)$. Onset of slow roll inflation is taken to
commence a little before the $k_o$ exits its Hubble horizon. The
value of the Hubble parameter at this time is so low that the total
energy density is less than $10^{-11}\rho_{\rm Pl}$. Therefore
throughout the inflationary era general relativity is an excellent
approximation to LQC. Equations of general relativity (or, LQC)
determine the mass $m$ of the inflaton as well as values of the
inflaton $\phi$ at $\t\eta(k_\star)$:
\be \label{m} m = 1.21 \times 10^{-6} \quad {\rm and} \quad
\phi(\t{t}(k_{\star})) = \pm 3.15\, .  \ee
Because of the observational error bars, these quantities are
uncertain by about 2\%. In the numerical simulations we use the
value of $m$ given in (\ref{m}). (For details, see \cite{as3}).

\bu \emph{Evolution of the Background:} So far numerical
evolutions of the background wave function $\Psi_o$ are
feasible only for kinetic dominated bounces, i.e., bounces for
which $\phi_{\B}$ is small. This is because the required time
over which one has to integrate to arrive in the general
relativity regime increases rapidly with $\phi_{\B}$.
Fortunately, as we will see below, this is the most interesting
portion of the allowed values of $\phi_{\B}$. These simulations
show that $\Psi_o$ remains sharply peaked on an effective
trajectory \cite{aps4}. Since there is no obvious reason why
this should not continue for higher $\phi_B$ values, it is
instructive to examine all effective trajectories without
restricting ourselves to kinetic energy dominated bounces. The
trajectory would be compatible with the 7 year WMAP data
\emph{only if at the point at which $H$ takes the value $7.83
\times 10^{-6}$, within the margin given by observational
errors,  $\epsilon = 8\times 10^{-3}$, and $\phi = 3.15$}. A
surprising result is that this is in fact the case under a
\emph{very} mild condition: In the $\phi_{\B} \ge 0$ sector,
for example, we only need $\phi_{\B} \ge 0.93$ \cite{as3}. Note
that this result is stronger than the qualitative `attractor
behavior' of inflationary trajectories because it is
quantitative and tuned to the details of the WMAP observations.
(For details, see \cite{as3}).

To make contact with the WMAP observations, we need to find
$k_\star$ and the time $\t\eta(k_\star)$ at which the mode with
co-moving wave number $k_\star$ exits the Hubble horizon during
inflation. For this, it is simplest to fix the scale factor at the
bounce and we will choose the convention $a_{\B} = 1$. (Note that
this is very different from $a_{\rm today} =1$ often used in
cosmology.) Then, along each dynamical trajectory one locates the
point at which the Hubble parameter takes the value $H = 7.83 \times
10^{-6}$ (and makes sure that at this time $\epsilon$ and $\phi$ are
given by (\ref{H}) and (\ref{m}) within observational errors). One
calls the conformal time at which this occurs $\t\eta(k_\star)$ and
numerically calculates the scale factor $a(\t{\eta}(k_\star))$ at
this time. Then, the value of the co-moving momentum $k_\star$ of
this mode is determined by the fact that this mode exits the Hubble
radius at time $\t\eta(k_\star)$. Thus, one asks that the
\emph{physical} wave number of this mode should equal the Hubble
parameter: $k/a({\t\eta}(k_\star)) = H(\t\eta(k_\star))$. Table 1
shows the values of $k_\star$, the physical wave length of the mode
at the bounce time, the \emph{proper} time $\t{t}(k_\star)$ at which
the mode exits the Hubble horizon, and the number of e-foldings
between the bounce and time $\t{t}(k_\star)$ for a range of values
of $\phi_{\B}$ which turns out to be physically most interesting.
(For details, see \cite{aan3}).

\begin{figure}[]
 \begin{center}
  \includegraphics[scale=1]{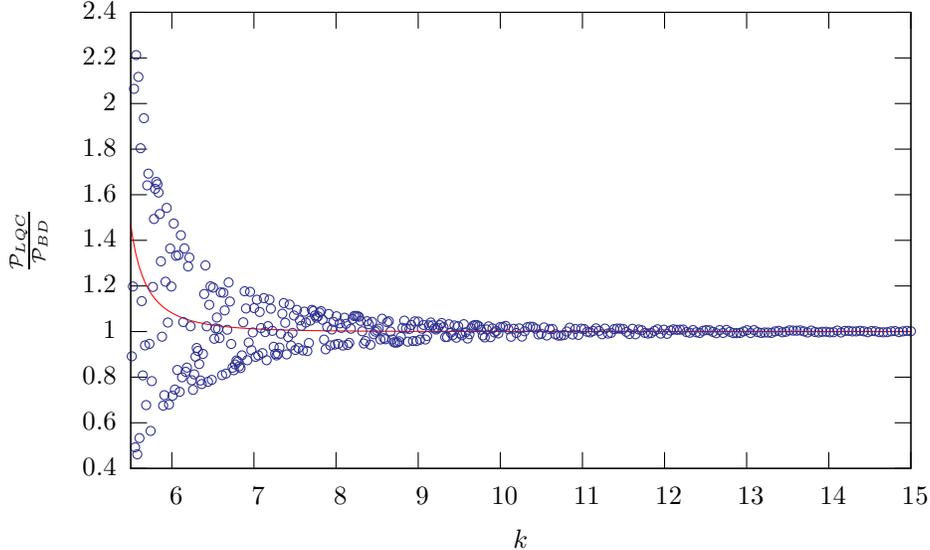}
  \caption{\label{fig3} Ratio of the LQC power spectrum for curvature
  perturbations in the scalar modes to that predicted by standard inflation
  (source \cite{aan3}). For small $k$, the ratio oscillates very rapidly.
  The (red) solid curve shows averages over (co-moving) bins with
  width $0.5\, {\lp}^{-1}$.}
 \end{center}
 \end{figure}

\bu \emph{Evolution of Perturbations:} Preliminary numerical
simulations were first carried out using four different states
$\psi$ at the bounce, satisfying the initial conditions discussed in
section \ref{s3}. They showed that the results are essentially
insensitive to the choice. Then detailed and much higher precision
simulations were carried out using $|\psi\rangle = |0_{\rm
obv}\rangle$, the `obvious vacuum of 4th adiabatic order', at the
bounce because, as mentioned before, this state has a number of
attractive properties. These simulations revealed an unforeseen
behavior: the power spectra for scalar and tensor perturbations are
largely insensitive to the value of $\phi_\B$. However, recall that
there is finite window $(k_o,\, 2000k_o)$\, of co-moving modes that
can be seen in the CMB. Because of the pre-inflationary dynamics,
the value of $k_\star$ ---and hence of $k_o$--- does depend on
$\phi_B$ and rapidly increases with $\phi_{\B}$. (See Table 1.)
Therefore, the window of observable modes \emph{is sensitive to the
value of $\phi_{\B}$ and moves steadily to the right as $\phi_{\B}$
increases.}

Fig.~\ref{fig3} shows the plot of the ratio $\mathcal{P}_{\R}^{\rm
LQC}/\mathcal{P}_{\R}^{\rm BD}$ of the LQC power spectrum to the
standard inflationary one for curvature perturbations $\R$ of the
scalar modes. The (blue) circles are the data points. The LQC power
spectrum has very rapid oscillations (whose amplitudes decay quickly
with $k$) which descend to the ratio that is plotted. Since
observations have only a finite resolution, to compare with data it
is simplest to average over small bins. We used bins which, at the
bounce, corresponds to a band-width in physical wave numbers of $0.5
t_{\rm Pl}^{-1}$. The result is the solid (red) line. We see that
the two power spectra agree for $k \gtrsim 6.5$ but LQC predicts an
enhancement for $k\lesssim 6.5$. We will now comment on these
features.

Let us first note that the LQC power spectrum in this plot uses the
value $\phi_{\B} = 1.15$. As Table 1 shows, the corresponding
$k_\star$ is $9.17$. At this value, the two power spectra are
identical, whence the amplitude and the spectral index obtained from
the LQC evolution at $k=k_\star$ agrees with the values
(\ref{observations}) observed by WMAP. However, as we remarked, for
$k \lesssim 6.5$, the LQC prediction departs from that of standard
inflation. These low $k$ values correspond to $\ell \lesssim 22$ in
the angular decomposition used by WMAP for which the error bars are
quite large. Therefore, although the LQG power spectrum differs from
the standard one in this range, both are admissible as far as the
current observations are concerned.

What is the physics behind the enhancement of the LQC power spectrum
for $k \lesssim 6.5$? And where does this specific scale come from?
This enhancement is due to pre-inflationary dynamics. At the bounce,
the scalar curvature has a universal value in LQC which sets a scale
$k_{\rm LQC} \approx 3.21$. Modes with $k \gg k_{\rm LQC}$
experience negligible curvature during their pre-inflationary
evolution while those with $k$ comparable to $k_{\rm LQC}$ or less
do experience curvature and therefore get excited. These are general
physical arguments and one needs numerical simulations to determine
exactly what `much greater than' and `comparable to' means. The
simulations show that modes with $k \gtrsim 2 k_{\rm LQC}$ already
satisfy the `much greater than' criteria. They are not excited and
for them the LQC state $\psi$ at the onset of inflation is virtually
indistinguishable from the BD vacuum. That is why the two power
spectra are essentially the same for $k \gtrsim 2 k_{\rm LQC}$. But
for modes with $k \lesssim 2k_{\rm LQC}$ the LQC state $\psi$ has
excitations over the BD vacuum whence there is an enhancement of the
power spectrum.

What happens if we change $\phi_{\B}$? As we remarked above, the
prediction of the LQC power spectrum is pretty insensitive to the
value of $\phi_{\B}$ but the window in the $k$ space spanned by
modes which are observable in the CMB changes, moving to the right
as $\phi_{\B}$ increases. Now, as Table 1 shows, if $\phi_{\B} >
1.2,$ we have $k_o > 6.5$, whence none of the observable modes would
be excited during the pre-inflationary evolution. In this case, at
the onset of the slow roll, the LQC sate $\psi$ would be
indistinguishable from the BD vacuum, whence all LQC predictions
would agree with those of standard inflation. Thus, there is a
narrow window, $0.93 \le \phi_{\B} \le 1.2$ for which the background
$\Psi_o$ admits the desired  slow roll phase and yet LQC predictions
for future observations can differ from the standard ones. One
example is given by a consistency relation $r = -8n_t$ in standard
inflation, where $r = 2\mathcal{P}_{\T}^{\rm BD}/\mathcal{P}_{\R}$
is the tensor to scalar ratio and $n_t$ is the spectral index for
tensor modes. This relation is significant because it does not
depend on the form of inflationary potential. It turns out that $r$
does not change in LQC but $n_t$ does, whence this standard
consistency relation is modified. Future observations would be able
to test for such departures. There is also a systematic study of the
effect that excitations over the BD vacuum can have on
non-Gaussianities
\cite{holman-tolley,agullo-parker,ganc,agullo-navarro-salas-parker}.
Furthermore, it has been recently pointed out that these
non-Gaussianities could be seen in the galaxy correlation functions
and also in certain distortions in the CMB
\cite{halo-bias1,halo-bias2,halo-bias3}. Thus, there are concrete
directions in which cosmological observations could soon start
probing effects that originate at the Planck scale. (For further
details, see \cite{aan3}).

\begin{figure}
 \begin{center}
  \includegraphics[scale=1]{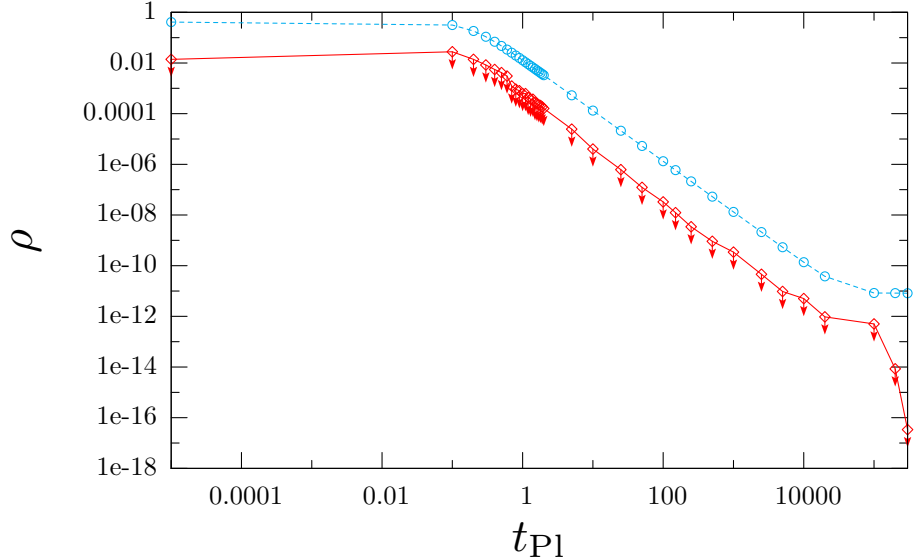}
  \caption{\label{fig4} For $\phi_{\B} =1.23 \mpl$, energy
  density in the background (upper curve) and an \emph{upper bound} on
  the energy density in perturbations (lower curve) are plotted
  against time from the bounce to the onset of slow roll using
  Planck units (source \cite{aan1}). The test field approximation
  holds across a change of 11 orders of magnitude in both quantities.}
 \end{center}
\end{figure}

\bu \emph{Self Consistency:} Finally, let us discuss the issue of
self-consistency of the truncation scheme, i.e., the issue of
whether the test field approximation continues to hold under
evolution. This issue is quite intricate and had remained unexplored
because of two different issues. The first issue is conceptual: it
was not clear how to compute the renormalized energy density for the
quantum fields $\h{\Q},\, \h{\T}$ in a manner that is meaningful in
the Planck regime. As discussed in section \ref{s5}, we were able to
construct this framework by `lifting' the adiabatic renormalization
theory on classical cosmological space-times to that on quantum
geometries $\Psi_o$. The second set of difficulties comes from
numerics: one requires very high accuracy and numerical precision.
This is because  i) the rapid oscillations of integrand of $\langle
\psi|\, \h\rho\, |\psi\rangle_{\rm ren}$ in the $k$ space make it
difficult to evaluate the exact value of the renormalized energy
density; and, ii) the background energy density itself decreases
from Planck scale to $10^{-11}$ times that scale. Indeed, so far we
have only managed to find an upper bound on the energy density in
the perturbations, shown in Fig.~\ref{fig4}. But this suffices to
show that, for $\phi_B > 1.22$, our initial conditions at the bounce
do give rise to a self-consistent solution $\Psi_o\otimes \psi$
throughout the evolution from the big bounce to the onset of slow
roll. These solutions provide a viable extension of the standard
inflationary scenario all the way to the Planck scale. The issue of
whether one can push the value of $\phi_{\B}$ to include the
interesting domain $\phi_{\B} < 1.2$ is still under investigation.
(There are several aspects to this problem, including a better
handling of the infrared regime, briefly discussed in \cite{aan3}.)

\section{Summary and Discussion}
\label{s7}

I began in section \ref{s1} by making some suggestions: i) Progress
in quantum gravity should be gauged by the degree to which an
approach succeeds in overcoming limitations of general relativity;
ii) The development of quantum theory, rather than general
relativity, offers a better example to emulate in this endeavor;
and, iii) As in quantum theory, it may be more fruitful to resolve
concrete physical problems at the interface of gravity and quantum
theory rather than focusing all efforts on obtaining a complete
quantum gravity theory in one stroke. In sections \ref{s2} and
\ref{s3} we saw that the very early universe offers an obvious arena
for this task for both conceptual and practical reasons.
Conceptually, the big bang is a prediction of general relativity in
a regime in which the theory is not applicable, whence it is
important to find out what really happened in the Planck regime. In
practical terms, currently the early universe offers the best hope
to confront quantum gravity theories with observations. In
particular, we saw that the inflationary paradigm has been highly
successful in accounting for the inhomogeneities in the CMB ---and
hence accounting for the large scale structure of the universe---
but it has several limitations. In sections \ref{s4} - \ref{s6}, I
summarized how the limitations related to the Planck scale physics
are being addressed in LQG. Specifically, by using the truncation
strategy of LQG, over the last six years it has been possible to
extend the inflationary paradigm all the way to the deep Planck
regime. (For other treatments of pre-inflationary dynamics within
LQG, see e.g. \cite{lqc-preinflation1,lqc-preinflation2}.)

The first finding is that the big bang singularity is resolved in
LQC and replaced by the big bounce. Since quantum physics
---including quantum geometry--- is regular at the big bounce, it is
natural to specify initial conditions for the quantum state $\Psi_o$
that encodes the background, homogeneous quantum geometry, as well
as for $\psi$ that describes the quantum state of perturbations.
Physically, the initial conditions amount to assuming that the state
$\Psi_o\otimes\psi$ at the bounce should satisfy `quantum
homogeneity'. More precisely, at the bounce one focuses just on that
region which expands to become the observable universe and demands
that it be homogeneous except for the inevitable quantum
fluctuations that one cannot get rid of even in principle. Now,
because of the pre-inflationary and inflationary expansion, the
region of interest has a radius smaller than $\sim 10 \lp$ at the
bounce. But as has been emphasized in the relativity literature,
this creates a huge fine tuning problem. For, to account for the
impressive fact that inhomogeneities in the CMB are really tiny
---just one part in $10^5$--- the required homogeneity at the bounce
has to be extraordinary. The standard inflationary paradigm is
not really applicable at the Planck scale and, even if one were
to ignore this fact, it does not have a natural mechanism to
achieve this degree of homogeneity. In LQC, on the other hand,
the big bang singularity is resolved precisely because there is
an in-built repulsive force with its origin in the specific
quantum geometry that underlies LQG. While this force is
negligible when curvature is less than, say, $10^{-6}$ in
Planck units, it rises spectacularly in the Planck regime,
overcomes the huge classical gravitational attraction and
prevents the big bang singularity. In more general models
referred to in section \ref{s4}, one finds a pattern: every
time a curvature scalar enters the Planck regime, this
repulsive force becomes dominant and dilutes that curvature
scalar, preventing a singularity (see e.g. \cite{asrev}). This
opens the possibility that the `dilution effect' of the
repulsive force may be sufficient to create the required degree
of homogeneity on the scale of about $10 \lp$, thereby
accounting for the assumed `quantum homogeneity'. If this idea
could to be developed in detail, dynamics of the pre-bounce
universe will leave no observable effects, providing a
clear-cut case for specifying initial conditions at the bounce.
Of course, the pre-bounce dynamics will still lead to
inhomogeneities at larger scales on the bounce surface but they
would have Fourier modes whose physical wave length is much
larger than the radius of the observable universe. Therefore,
they would not be in the observable range; in the truncated
theory considered here, they would be absorbed in the quantum
geometry of the homogeneous background. This `dilution
mechanism' and other issues related to initial conditions are
likely to be a center of activity in the coming years.

As we saw in sections \ref{s5} and \ref{s6}, we now have a
conceptual framework and numerical tools to evolve these initial
conditions all the way from the bounce to the onset of slow roll.
The result depends on where one is in the parameter space that is
labeled by the value $\phi_{\B}$ of the inflaton at the bounce. For
a very large portion of the parameter space we obtain the following
three features: i) Some time in its future evolution, the background
geometry encounters a slow roll phase that is compatible with the 7
year WMAP observations;  ii) At the onset of this slow roll, the
state $\psi$ of perturbations is essentially indistinguishable from
the BD vacuum used in standard inflation; and iii) the back reaction
due to perturbations remains negligible throughout pre-inflationary
dynamics in which the background curvature falls by some 11 orders
of magnitude, justifying the underlying `truncation approximation'.
Thus, for this portion of the parameter space, we have a
self-consistent extension of the standard inflationary paradigm.

There is, however, a small window in the parameter space for which
the feature i) is realized but the initial state at the onset of
inflation contains an appreciable number of BD excitations. This
number is within the current observational limits. But the presence
of these excitations signals new effects such as a departure from
the inflationary `consistency relation' involving both scalar and
tensor modes and a new source of non-Gaussianities. These could be
seen in future observational missions
\cite{halo-bias1,halo-bias2,halo-bias3}. The physical origin of
these effects can be traced back to a new energy scale $k_{\rm LQC}$
defined by the universal value of the scalar curvature at the
bounce. Excitations with $k \lesssim 2k_{\LQC}$ are created in the
Planck regime near the bounce. It turns out that if the number $N$
of e-foldings in the scale factor $a$ between the bounce and $\t\eta
= \t\eta_{k^\star}$ is less than 15, then the modes which are
excited would be seen in the CMB. This occurs only in the small
window of parameter space referred to above. Since the window is
very small, the `a priori probability' that one of these values of
$\phi_{\B}$ is realized in Nature would seem to be tiny. However,
one can turn this argument around. Should these effects be seen, the
parameter space would be narrowed down so much that very detailed
calculations would become feasible. In either case, it is rather
exciting that the analysis relates initial conditions and Planck
scale dynamics with observations, thereby expanding the reach of
cosmology to the earliest moment in the deep Planck regime.

Even when a self-consistent solution $\Psi_o\otimes\psi$ to the
truncated theory exists, how would it fit in full LQG? Recall the
situation in classical general relativity. In cosmology as well as
black hole physics, one routinely expects first order perturbations
whose back reaction is negligible to provide excellent
approximations to the phenomenological predictions of the exact
theory. I see no obvious reason why the situation would be different
in quantum gravity. As a simple example to illustrate the general
viewpoint, consider the Dirac solution of the hydrogen atom. Since
one assumes spherical symmetry prior to quantization, this
truncation excludes photons from the beginning. Therefore, at a
conceptual level, the Dirac description is \emph{very} incomplete.
Yet, as far as experiments are concerned, it provides excellent
approximations to answers provided by full QED until one achieves
the accuracy needed to detect the Lamb shift. I expect the situation
to be similar for our truncated theory: Conceptually it is surely
quite incomplete vis a vis full LQG, but the full theory will
provide only small corrections to the observable effects.

To conclude, let me emphasize that there was no a priori reason to
anticipate either of the two main conclusions ---the extension of
standard inflation to the Planck regime for much of the parameter
space and deviations from some of its predictions in a narrow
window. Indeed, it would \emph{not} have been surprising if the
pre-inflationary dynamics of LQC was such that the predicted power
spectra were observationally ruled out for the `natural' initial
conditions we used at the bounce, or, if the self-consistency of
truncation had failed quite generally because of the Planck scale
dynamics. Indeed, this could well occur in generic bouncing
scenarios, e.g. in situations in which the expansion between the
bounce and the surface of large scattering is not sufficiently large
for the modes observed in the CMB to have wave lengths smaller than
the curvature radius throughout this evolution.

\section*{Acknowledgments}
This report is based  on joint work with Ivan Agullo, Alejandro
Corichi,  Wojciech Kaminski, Jerzy Lewandowski, William Nelson,
Tomasz Pawlowski, Parampreet Singh and David Sloan over the past six
years. I am most grateful for this collaboration. I am also indebted
to a very large number of colleagues especially in the LQG community
for discussions, comments, questions and criticisms. This work was
supported by the NSF grant PHY-1205388 and the Eberly research funds
of Penn state.

\end{document}